# OpenRISC System-on-Chip Design Emulation

Kai Cong, Li Lei, and Zhenkun Yang
Advisor: Fei Xie
{congkai, leil, zhenkun, xie}@cs.pdx.edu
Department of Computer Science, Portland State University, Portland, OR

## 1. Introduction

New computer systems like smartphones and tablets, are entering the market at an ever-accelerating pace. This brings enormous pressure on the product development teams to shorten the *time-to-market*. Driven by increasing design complexity and decreasing time-to-market, it demands innovative approaches to accelerating hardware design simulation, verification, and debugging. Recently the hardware emulation technique has emerged as a promising approach to accelerating hardware verification/debugging process.

To fully evaluate the powerfulness of the emulation approach and demonstrate its potential impact, we propose to emulate a system-on-chip (SoC) design using Mentor Graphics Veloce emulation platform. This article presents our project setup and the results we have achieved. In this project, we carry out the following tasks: (1) standalone emulation of an existing open-source SoC design, **O**penRISC **R**eference **P**latform **S**ystem **o**n **C**hip (ORPSoC); (2) emulation performance evaluation with three categories of benchmarks, running 'sum' program with different parameters over ORPSoC, booting the Linux kernel over ORPSoC and running a set of software programs over ORPSoC; (3) thoroughly comparison with the simulation approach: simulating ORPSoC with the benchmarks on Mentor Graphics ModelSim.

The results are encouraging. ORPSoC emulation with Veloce has more than ten times faster than hardware simulation. Our experimental results demonstrate that Mentor Graphics Veloce has major advantages in emulation, verification, and debugging of complicated real hardware designs, especially in the context of SoC complexity. Through our three major tasks, we will demonstrate that (1) Veloce can successfully emulate large-scale SoC designs; (2) it has much better performance comparing to the state-of-the-art simulation tools; (3) it can significantly accelerate the process of hardware verification and debugging while maintaining full signal visibility.

## 2. Background

**OpenRISC 1200:** OpenRISC 1200 (OR1200) [1] is a synthesizable CPU core developed and maintained by developers at OpenCores [2]. The OR1200 design is an open source implementation of the OpenRISC 1000 RISC architecture [3], which is implemented in Verilog HDL. OR1200 has following major features:
- Central CPU/DSP block
- Direct mapped data/instruction cache
- WISHBONE bus interfaces

**ORPSoC**: ORPSoC is an OpenRISC-based reference SoC [4], which consists of following hardware components, as shown in Figure 1.
- WISHBONE Bus
- SRAM Memory
- General-purpose I/O (GPIO)
- UART serial port
- 10/100Mbps Ethernet MAC

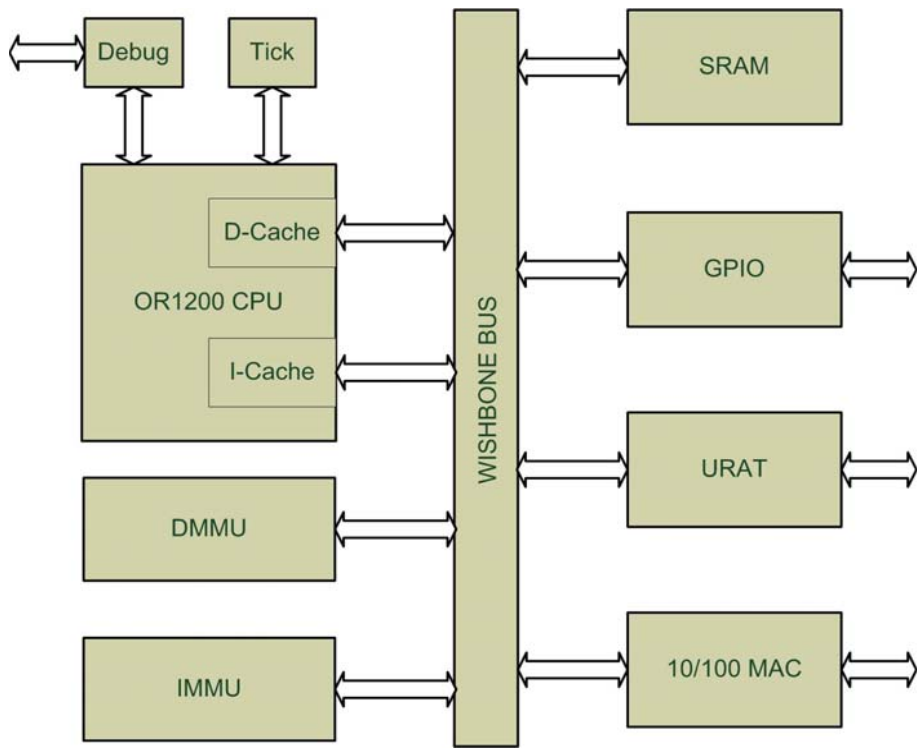

Figure-1: Architecture of ORPSoC with major components

## 3. ORPSoC Simulation

We use Mentor Graphics ModelSim as our RTL design simulator. The workflow of our simulation is depicted as Figure-2.

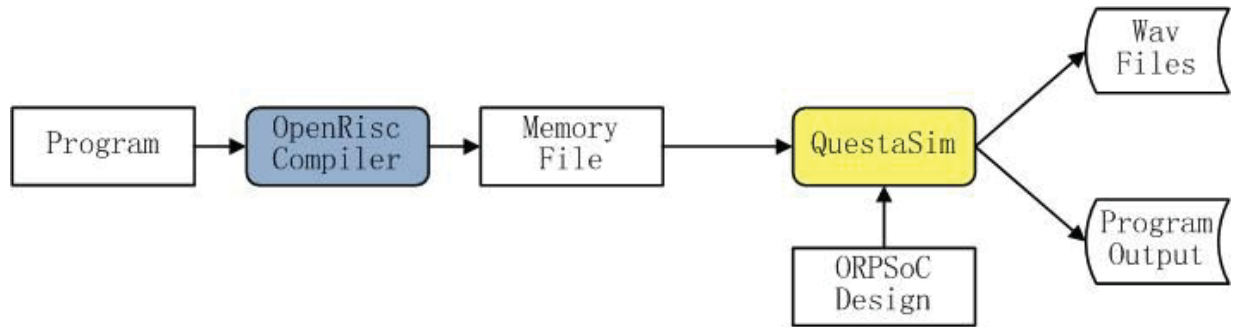

Figure-2: Workflow of ORPSoC Simulation with ModelSim

The workflow contains three steps:
1). By using the OpenRISC compiler, the target program running on ORPSoC is compiled into a memory file which is consumed by the simulator ModelSim.
2). ModelSim takes the compiled ORPSoC design and the memory file. It simulates the target program running on the ORPSoC design.
3). ModelSim stops when the program terminates. It outputs the waveform of the selected signals of the ORPSoC design and records the program output in the log file.

## 4. ORPSoC Emulation

### 4.1 Workflow of ORPSoC Emulation.

The workflow of our ORPSoC emulation with Mentor Graphics Veloce is illustrated as Figure-3. In this workflow, the memory file generated from the target program plays as the testbench which includes a significant amount of CPU instructions on our ORPSoC design.

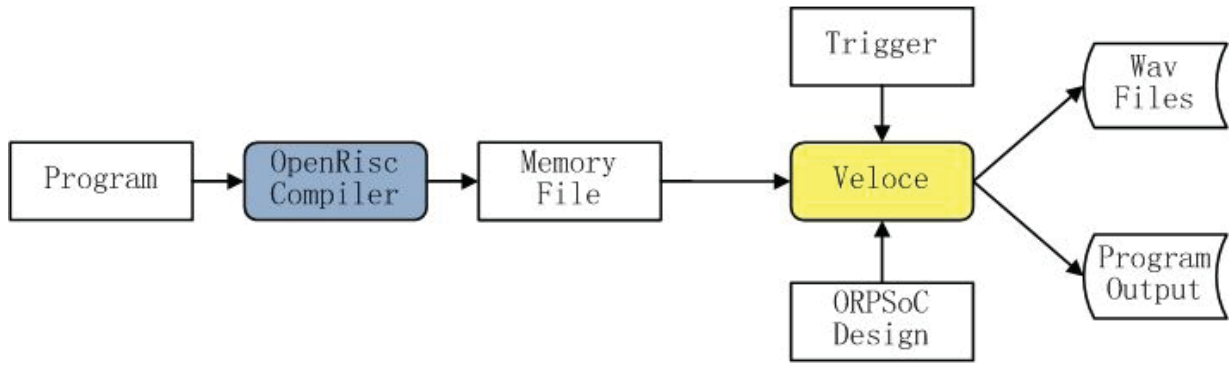

Figure-3: Workflow of ORPSoC emulation with Veloce

The workflow contains four steps:
1). Using the OpenRISC compiler, the target program running on ORPSoc design is compiled into a memory file which is loaded into the memory.
2). Veloce software first compiles and synthesizes the ORPSoc design. Then the design is configured on the Veloce machine.
3). In order to terminate the emulation automatically, triggers are defined and applied to the emulation process.
4). Before the emulation starts, both memory and triggers are downloaded into Veloce machine. Then emulation is executed and wave files and related information are generated after the emulation is terminated by the trigger.

### 4.2 Emulation Triggers

Veloce provides trigger mechanism which allows detection of a given state of logic during emulation. In our project, we mainly define two triggers for Linux System boot-up and generic programs separately. Both triggers are used for determining the termination condition during emulation. Using the Veloce trigger editor, triggers are created and trigger diagrams are generated easily.

### 4.2.1 Triggers for Linux System boot-up

For Linux system boot-up, we need to determine when the system is booted. We first instrumented the linux kernel source code to output some special flag through uart module. Then we define a trigger to capture such special flag to notify that the system has booted. The corresponding trigger diagram is as follows:

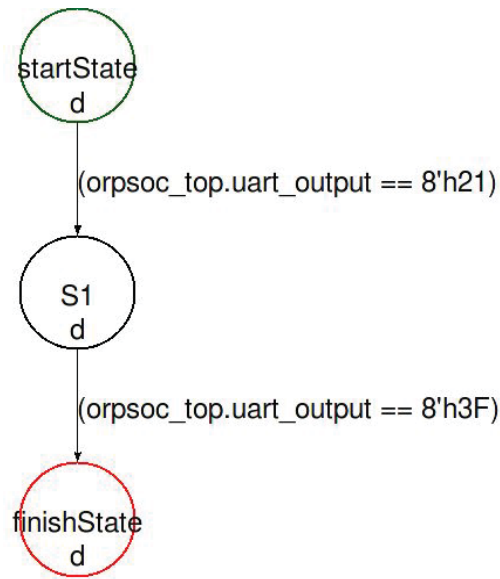

### 4.2.2 Triggers for generic programs

For the generic programs including CHStone benchmark and 'SUM' program, we need to determine when the program has been executed. We added a special instruction into the end of the program memory file. Then we define a trigger to capture such special instruction to notify that the program has been executed. The corresponding trigger diagram is as follows:

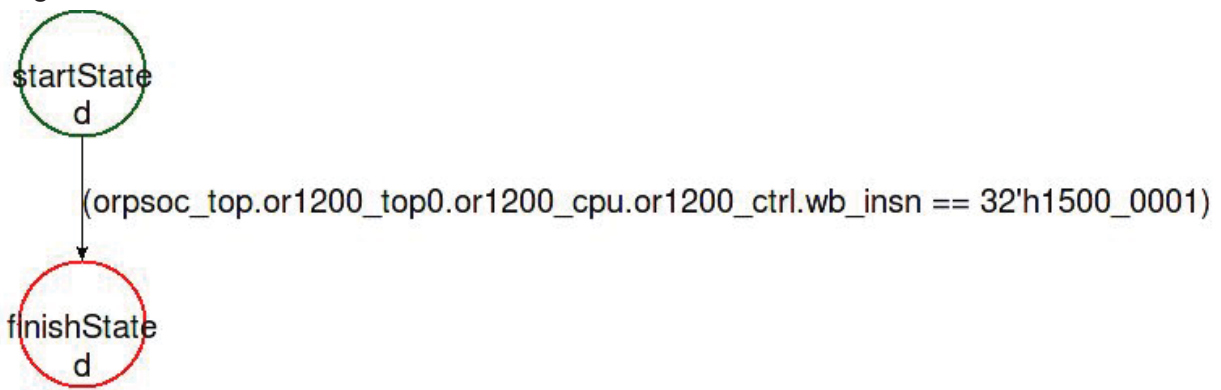

## 5. Evaluation

We mainly focus on evaluating how much speedup can be achieved when Veloce deals with a complicated hardware design. In this section, we evaluate the performance of two approaches:
- ORPSoC Simulation with ModelSim

- ORPSoC Emulation with Veloce

Moreover, we evaluate the two approaches under three benchmarks:
- Sum N
- Linux System boot-up
- CHStone benchmark, which consists of 12 designs

### 5.1 ORPSoC Information

Table-I shows basic information of the ORPSoC design. Blank, comment, and code are evaluated in Lines of Code (LOC).

Table-I: ORPSoC RTL Design Information

| Language | Files | Blank | Comment | Code |
|---|---|---|---|---|
| Verilog | 201 | 7466 | 21276 | 46611 |

### 5.2 Evaluation Benchmark 1: Sum N

We evaluate simulation and emulation of the following 'sum' program with different *N* with ModelSim and Veloce. From Table-II and Figure-4, we can see that emulation has major advantage over simulation when executing large number of instructions. The results demonstrate that (1) when the number of iterations is small, there is no significant speedup achieved by Veloce; (2) As the number of iterations increases, the speedup achieved by Veloce is significantly improved.

```c
#define N 10
int main() {
    long int i = 0, sum = 0;
    for (i = 0; i < N; i++)
        sum += i;
    return 0;
}
```

Table-II: Evaluation Results of Sum with Different N

| N | Simulation Time (s) | Emulation Time (s) | Speedup (x) |
|---|---|---|---|
| 10 | 1.908 | 1.663 | 1.15 |
| 100 | 1.982 | 1.7 | 1.17 |
| 1000 | 3.122 | 1.66 | 1.88 |

| 10,000 | 13.776 | 1.788 | 7.70 |
| 100,000 | 120.987 | 3.356 | 36.05 |
| 1,000,000 | 1216.704 | 18.941 | 64.24 |
| 10,000,000 | 13011.505 | 175.612 | 74.09 |

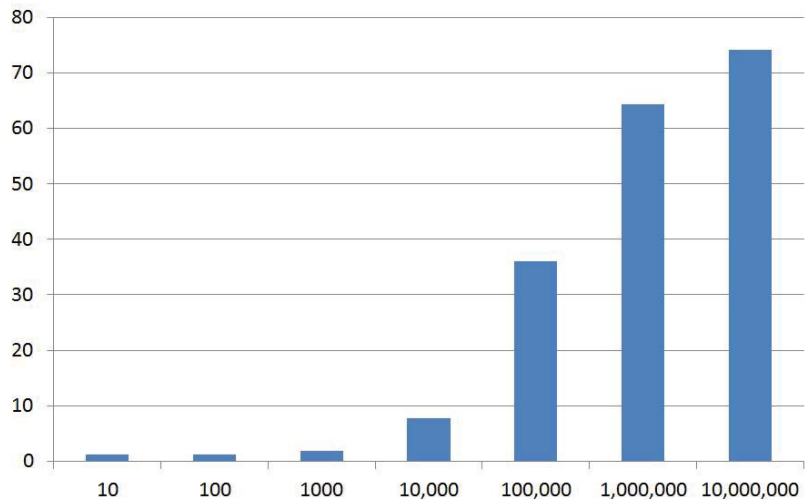

Figure-4 Speedup achieved by Veloce with N increases

## 5.3 Evaluation Benchmark 2: Booting Linux System on ORPSoC

We evaluate the performance of booting a Linux system on ORPSoC. We compile the Linux kernel from the source, and build the vmem file from the compiled binary. We simulated/emulated booting a Linux system on the CPU. It takes about 55 million of instruction and 198 million of clock cycles to boot the Linux. Veloce takes about 127s and Xilinx Spartan-6 FPGA with 50MHz clock takes about 30s to boot the Linux system. Theoretically, FPGA should take shorter time, it took longer than expected is because it need to program the FPGA from FLASH memory on boot. Table-III shows the statistics of booting Linux system. However, we didn't get the time from ModelSim, it got stuck during the simulation. We timed it at the beginning, ModelSim took about ~2s to run 10,000 instructions, theoretically ModelSim should take about 3 hours to boot Linux.

Running an operating system is an important test scenario while developing and testing SoC designs. In practice, the simulation approach is often not acceptable as it takes hours to just boot the system, let alone debugging the software running in the system. Veloce not only provides the designers full traceability over entire software stacks and hardware systems, but also enables significant speedup over the simulation approach.

Our evaluation on booting a Linux system on ORPSoC fully demonstrates the benefits Veloce brings us.

Table-III: CHStone Benchmark Evaluation Results

|  | # of Instructions | # cycles | ModelSim Simulation | Veloce Emulation | FPGA (50MHz) |
|---|---|---|---|---|---|
| Linux boot-up | ~55 million | ~198 million | ~3 hours | ~127.1s | ~30s |

## 5.4 Evaluation Benchmark 3: CHStone

CHStone [5] is a C-based benchmark which consists of 12 programs which are selected from various application domains such as arithmetic, media processing, security and microprocessor. Table-IV shows the complexity of the benchmark. We compiled all the designs into executable binary files, and then convert the binary files into vmem files. Table-IV and Figure-5 demonstrate that (1) when running the applications with the low complexity, such as GSM, the speedup achieved by Veloce is not obvious; (2) while running the applications with the high complexity, such as DFSIN, Veloce achieved significant speedup.

This designs in CHStone benchmark are real industry applications which are widely used. Running these applications on ORPSoC with simulation and emulation demonstrates that Veloce also provides benefits in testing SoC designs at application-level.

Table-IV: CHStone Benchmark Evaluation Results

| App. Domain | Design | LoC | # of Instructions | Simulation Time (s) | Emulation Time (s) | Speed-up |
|---|---|---|---|---|---|---|
| Arithmetic | DFADD | 526 | 8943366 | 248.272 | 14.428 | 17.21 |
|  | DFDIV | 436 | 4328401 | 116.053 | 7.812 | 14.86 |
|  | DFMUL | 376 | 3909483 | 104.182 | 7.228 | 14.41 |
|  | DFSIN | 755 | 13274785 | 335.227 | 19.366 | 17.31 |
| Microprocessor | MIPS | 232 | 64311 | 3.136 | 1.704 | 1.84 |
| Media Proc. | ADPCM | 541 | 145840 | 5.509 | 1.812 | 3.04 |
|  | GSM | 393 | 25776 | 2.275 | 1.668 | 1.36 |
|  | JPEG | 1692 | 4957468 | 137.109 | 8.135 | 16.85 |
|  | MOTION | 583 | 62126 | 3.107 | 1.699 | 1.83 |
| Security | AES | 716 | 275207 | 9.102 | 2.009 | 4.53 |
|  | BLOWFISH | 1406 | 1449807 | 37.703 | 1.812 | 20.81 |
|  | SHA | 1284 | 844421 | 23.792 | 2.358 | 10.09 |

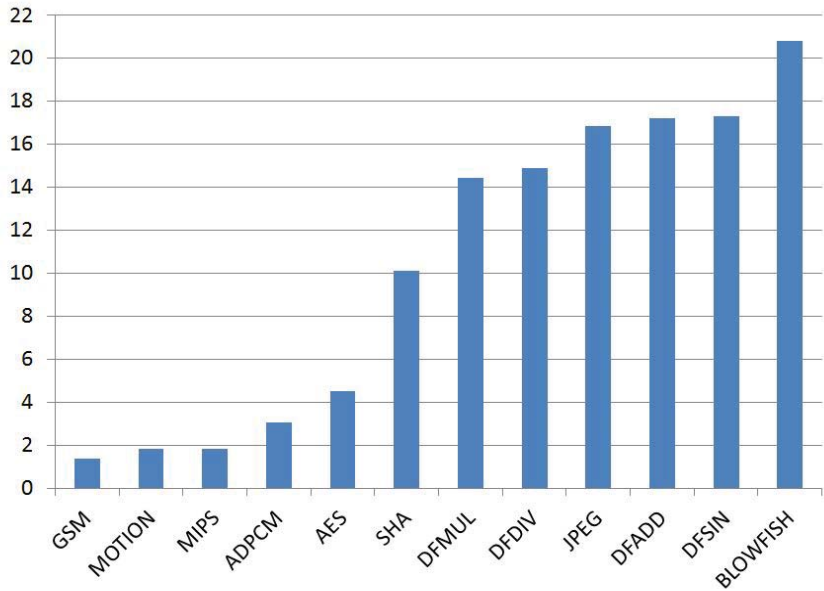

Figure-5 Speedup achieved by Veloce on CHStone benchmark

## 6. Discussion and Research in Our Group

Nowadays there are so many electronic products entering the market everyday. To earn more market share, it requires electronic companies to shorten the product development cycles and deliver high-quality products. Both academic research groups and electronic companies are exploring innovative and systematic approaches to shorten the time-to-market, reduce the development cost and improve the product quality. In our group, we have been conducting several critical research for achieving this goal in the past several years.

1). ***Post-silicon functional validation with virtual prototypes.*** Post-silicon validation has become a critical stage in the system-on-chip (SoC) development cycle, driven by increasing design complexity, higher level of integration and decreasing time-to-market. According to recent reports, post-silicon validation effort comprises more than 50% of the overall development effort of an 65nm SoC. Though post-silicon validation covers many aspects ranging from electronic properties of hardware to performance and power consumption of whole systems, a central task remains validating functional correctness of both hardware and its integration with software. There are several key challenges to achieving accelerated and low-cost post-silicon functional validation. First, there is only limited silicon observability and controllability; second, there is no good test coverage estimation over a silicon device; third, it is difficult to generate good post-silicon tests before a silicon device is available; fourth, there is no effective software robustness testing approaches to ensure the quality of hardware/software integration. We propose a systematic approach to accelerating post-silicon functional validation with virtual prototypes [6]. Post-silicon test coverage [7] is estimated in the pre-silicon stage by evaluating the test cases on the virtual prototypes. Such analysis is first conducted on the initial test suite assembled by the user and subsequently on the expanded test suite which includes test cases that are automatically generated [8]. Based on the coverage statistics of the initial test suite on the virtual prototypes, test cases are automatically generated to improve the test coverage using symbolic execution [9]. In the post-silicon stage, our approach supports coverage evaluation of test cases on silicon devices to ensure fidelity of early coverage evaluation. The generated test cases are issued to silicon devices to detect inconsistencies between virtual prototypes and silicon devices using conformance checking. We further extend the test case generation framework to generate and inject fault scenario with virtual prototypes for driver robustness testing. Besides virtual prototype-based fault injection, an automatic driver fault injection approach is developed to support runtime fault generation and injection for driver robustness testing [10]. Since virtual prototype enables early driver development, our automatic driver fault injection approach can be applied to driver testing in both pre-silicon and post-silicon stages. For preliminary evaluation, we have applied our coverage evaluation and test generation to several network adapters and their virtual prototypes. We have conducted coverage analysis for a suite of common tests on both the virtual prototypes and silicon devices. The results show that our approach can estimate the test coverage with high fidelity. Based on the coverage estimation, we have employed our automatic test generation approach to generate additional tests. When the generated test cases were issued to both virtual prototypes and silicon devices, we observed significant coverage improvement. And we detected 20 inconsistencies between virtual prototypes and silicon devices, each of which reveals a virtual prototype or silicon device defect. After we applied virtual prototype-based fault injection approach to virtual prototypes for three widely-used network adapters, we generated and injected thousands of fault scenarios and found 2 driver bugs. For automatic driver fault injection, we have applied our approach to 12 widely used drivers with either virtual prototypes or silicon devices. After testing all these drivers, we found 28 distinct bugs.

2). ***Hardware/Software Interface Assurance with Conformance Checking.*** Hardware/Software (HW/SW) interfaces are pervasive in modern computer systems. Most of HW/SW interfaces are implemented by devices and their device drivers. Unfortunately, HW/SW interfaces are unreliable and insecure due to their intrinsic complexity and error-prone nature. Moreover, assuring HW/SW interface reliability and security is challenging. First, at the post-silicon validation stage, HW/SW integration validation is largely an ad-hoc and time-consuming process. Second, at the system deployment stage, transient hardware failures and malicious attacks make HW/SW interfaces vulnerable even after intensive testing and validation. In this dissertation, we present a comprehensive solution for HW/SW interface assurance over the system life cycle. This solution is composed of two major parts. First, our solution provides a systematic HW/SW co-validation framework which validates hardware and software together; Second, based on the co-validation framework, we design two schemes for assuring HW/SW interfaces over the system life cycle: (1) post-silicon HW/SW co-validation at the post-silicon validation stage; (2) HW/SW co-monitoring at the system deployment stage. Our HW/SW co-validation framework employs a key technique, conformance checking which checks the interface conformance between the device and its reference model [11] [12]. Furthermore, property checking is carried out to verify system properties over the interactions between the reference model and the driver [13]. Based on ii the conformance between the reference model and the device, properties hold on the reference model/driver interface also hold on the device/driver interface.

Conformance checking discovers inconsistencies between the device and its reference model thereby validating device interface implementations of both sides. Property checking detects both device and driver violations of HW/SW interface protocols. By detecting device and driver errors, our co-validation approach provides a systematic and efficient way to validate HW/SW interfaces [14]. We developed two software tools which implement the two assurance schemes: DCC (Device Conformance Checker), a co-validation framework for post-silicon HW/SW integration validation; and CoMon (HW/SW Co-monitoring), a runtime verification framework for detecting bugs and malicious attacks across HW/SW interfaces. The two software tools lead to discovery of 42 bugs from four industry hardware devices, the device drivers, and their reference models. The results have demonstrated the significance of our approach in HW/SW interface assurance of industry applications.

3). ***Scalable Equivalence Checking for Behavioral Synthesis.*** Behavioral synthesis is the process of compiling an Electronic System Level (ESL) design to a register-transfer level (RTL) implementation. ESL specifications define the design functionality at a high level of abstraction (e.g., with C/C++ or SystemC), and thus provide a promising approach to address the exacting demands to develop feature-rich, optimized, and complex hardware systems within aggressive time-to-market schedules. Behavioral synthesis entails application of complex and error-prone transformations during the compilation process. Therefore, the adoption of behavioral synthesis highly depends on our ability to ensure that the synthesized RTL conforms to the ESL description. This research provides an end-to-end scalable equivalence checking support for behavioral synthesis. The major challenge of this research is to bridge the huge semantic gap between the ESL and RTL descriptions, which makes the direct comparison of designs in ESL and RTL difficult. Moreover, a large number and a wide variety of aggressive transformations from front-end to back-end require an end-to-end scalable checking framework. A behavioral synthesis flow can be divided into three major phases, including 1) *front-end*: compiler transformations, 2) *scheduling*: assigning each operation a clock cycle and satisfying the user-specified constraints, and 3) *back-end*: local optimizations and RTL generation. In our end-to-end and incremental equivalence checking framework, we check each of the three phases one by one. Firstly, we check the front-end that consists of a sequence of compiler transformations by decomposing it into a series of checks, one for each transformation applied [16]. We symbolically explore paths in the input and output programs of each transformation, and check whether the input and output programs have the same observable behavior under the same path condition. Secondly, we validate the scheduling transformation by checking the preservation of control and data dependencies, and the preservation of I/O timing in the user-specified scheduling mode[17]. Thirdly, we symbolically simulate the scheduled design and the generated RTL cycle by cycle, and check the equivalence of each mapped variables. We also develop several key optimizations to make our back-end checker scale to real industrial-strength designs [15]. In addition to the equivalence checking framework, we also present an approach to detecting deadlocks introduced by parallelization of RTL blocks that are connected by synthesized interfaces with handshaking protocols.

## 7. Conclusion and Future Work

In this project, we simulate and emulate an existing open-source SoC design based on OpenRISC architecture. We successfully simulated the design with ModelSim. More importantly, we port the design to Veloce emulation platform, and emulated the whole SoC design. We evaluated the performance of simulation versus emulation on two benchmarks: Linux System boot-up and CHStone. The results demonstrate that when ORPSoC executes a large number of instructions, the simulation speed is significant slow, which may not be acceptable in hardware verification and debugging. With the help of Veloce, we achieved significant speedup over ModelSim. As our non-trivial evaluation benchmarks are common SoC testing scenarios in industry, the results demonstrate that Veloce has a large potential to facilitate hardware verification/debugging in the real industry practice.

In the future, we plan to add more peripheral devices to the OpenRISC SoC and evaluate Veloce in depth. We also want to explore advanced features in Veloce, such as TBX. We plan to build a SoC platform with customized peripherals. Based on the platform, we will conduct research on SoC debugging and verification with Veloce.